\documentclass[twoside,onecolumn,9pt,notitlepage]{article}
\usepackage{extsizes}
\usepackage[super,sort&compress,comma]{natbib} 
\newcommand{\textcite}[1]{[\citenum{#1}]} 

\usepackage{amsmath}
\usepackage[left=1.5cm, right=1.5cm, top=1.785cm, bottom=2.0cm]{geometry}
\usepackage{balance}
\usepackage{mathptmx}
\usepackage{sectsty}
\usepackage{graphicx} 
\usepackage{lastpage}
\usepackage[format=plain,justification=justified,singlelinecheck=false,font={stretch=1.125,small,sf},labelfont=bf,labelsep=space]{caption}
\usepackage{float}
\usepackage{fancyhdr}
\usepackage{fnpos}
\usepackage[english]{babel}
\addto{\captionsenglish}{%
  
}
\usepackage{array}
\usepackage{droidsans}
\usepackage{charter}
\usepackage[T1]{fontenc}
\usepackage[usenames,dvipsnames]{xcolor}
\usepackage{setspace}
\usepackage[compact]{titlesec}
\usepackage[colorlinks=true,allcolors = {black}]{hyperref}

\usepackage{soul}
\usepackage{mathrsfs} 
\newcommand{\ldoc}{\mathscr{C}}
\newcommand{\ldocbar}{\overline{\mathscr{C}}}

\newcommand{\ldocoabar}{\langle\overline{\mathscr{C}}\rangle}
\newcommand{\tmatrix}{\emph{T}-matrix}

\definecolor{cream}{RGB}{222,217,201}

\usepackage{titling}
\usepackage{authblk}

\pretitle{\begin{center}\Huge\bfseries}
\posttitle{\par\end{center}\vskip 0.5em}
\preauthor{\begin{center}}
\postauthor{\end{center}}
\predate{\par\centering}
\postdate{\par}

\title{Orientation dependence of optical activity in light scattering by nanoparticle clusters}
\author[1,2]{\Large Atefeh Fazel-Najafabadi}
\author[1,2]{\Large Baptiste Augui\'e}
\affil[1]{\normalsize School of Chemical and Physical Sciences\protect\\
Victoria University of Wellington, PO Box 600,
  Wellington 6140, New Zealand}
\affil[2]{\normalsize The MacDiarmid Institute for
  Advanced Materials and Nanotechnology}
\date{\today}

\begin{document}

\twocolumn[
\begin{@twocolumnfalse}

\maketitle

\begin{abstract}
The optical properties of nanoparticle clusters vary with the spatial arrangement of the constituent particles, but also the overall orientation of the cluster with respect to the incident light. This is particularly important in the context of nanoscale chirality and associated chiroptical responses, such as circular dichroism or differential scattering of circularly polarised light in the far-field, or local degree of optical chirality in the near-field. We explore the angular dependence of such quantities for a few archetypal geometries: a dimer of gold nanorods, a helix of gold nanospheres, and a linear chain of silicon particles. The examples serve to illustrate the possible variation of chiroptical responses with the direction of light's incidence, but also, consequently, the importance of a robust orientation-averaging procedure when modelling general clusters of particles in random orientation. Our results are obtained with the rigorous superposition \tmatrix\ method, which provides analytical formulas for exact orientation-averaged properties.
\end{abstract}

\vskip 2em
\end{@twocolumnfalse}
]


\pagestyle{empty}
\thispagestyle{empty}
\fancypagestyle{plain}{
\renewcommand{\headrulewidth}{0pt}
}

\section{Introduction}
%

%

Light scattering by assemblies of nanoparticles underpins many important applications of advanced nanomaterials, notably in the realm of optical sensing\cite{mcfarland2003single, saylan2020plasmonic, kasani2019review, neubrech2020reconfigurable}, spectroscopy\cite{le2008principles, maccaferri2021recent, stamplecoskie2011optimal, haes2005plasmonic}, nano-optics\cite{novotny2012principles, lal2010nano, sandoghdar2020nano}, and light-harvesting technologies\cite{carretero2016plasmonic}. It is also key to their characterisation\cite{djorovic2020extinction}. Bottom-up synthesis and assembly of nanoparticles enables fine-tuned optical properties, where the particles or clusters of particles are often fabricated, characterised, and used in colloidal form. In later characterisation or application, the particles may be attached on a substrate and assume preferential orientations. Nanoparticles are often used in combinations; in fact, the assembly of nanoparticles into clusters of well-defined geometry is a powerful means to realise optical properties that cannot be achieved with single particles. For instance, one of the best-performing configurations for surface-enhanced spectroscopies is to form a very small gap between two or more particles, strongly enhancing local electric fields in the gap region\cite{le2008principles}. In the context of nanoscale chirality, the assembly of nanoparticles into three-dimensional clusters has proven in recent years a very fruitful strategy\cite{govorov2011chiral, guerrero2011individual, fan2013optical, lan2015nanorod}, notably to overcome the intrinsic difficulty in synthesising individual chiral nanoparticles\cite{govorov2011chiral, guerrero2011individual}. Some of the first structures to be proposed were dissymetric tetrahedra\cite{fan2013optical, yan2012self} and helices of nanospheres\cite{kuzyk2012dna, govorov2011chiral, shen2012rolling, jung2014chiral}. With nonspherical particles such as nanorods, only 2 particles are needed to produce a chiral "fingers crossed" assembly\cite{auguie2011fingers, yin2013interpreting, najafabadi2017analytical, lan2013bifacial, zhao2017chirality}, which has been realised in a variety of ways\cite{yin2015active, jiang2017stimulus, zhou2015plasmonic, lan2013bifacial, ma2013chiral, zhao2017chirality}, such as with DNA origami\cite{jiang2017stimulus, zhou2015plasmonic,lan2013bifacial}. The optical properties of nanoparticle clusters can differ dramatically from those of the constituent particles; this occurs when the presence of the neighbouring particles substantially modifies the net field exciting each particle. A typical example is a dimer, where the field scattered by one particle contributes to the net excitation of its neighbour, and vice versa\cite{yin2013interpreting, najafabadi2017analytical}. Such inter-particle coupling can occur cumulatively over several particles, provided the scattered fields add up with a well-defined phase relationship\cite{lasa2020surface}. This means that the precise geometrical arrangement of particles in a cluster and the direction and polarisation of the light wave exciting them lead to a delicate interaction\cite{hentschel2012three, bordley2015coupling, fan2013optical}, that is revealed in the overall optical properties. 
With chiral geometries, one is often concerned with the difference in optical response when the assembly is excited by circularly polarised light of either left or right handedness. In the far-field, this is measured as circular dichroism\cite{Berova:2012aa}, for light extinction, or differential scattering or absorption\cite{wang2015circular, ma2013attomolar}. The recent surge of interest in polarised light's interaction with chiral nanostructures has also turned to near-field properties, following the proposal by Cohen \emph{et al} \cite{TangCohen2010} of locally-enhanced degree of optical chirality, $\ldoc$, in the vicinity of suitably-engineered nanostructures. This quantity potentially leads to an enhancement of chiroptical spectroscopies of molecular species, or enormous practical importance for pharmaceutical and bio/chemical compounds\cite{severoni2020plasmon, garcia2019enhanced, ma2013attomolar}.

\begin{figure}[!htpb]
	\centering
\includegraphics[width=\columnwidth]{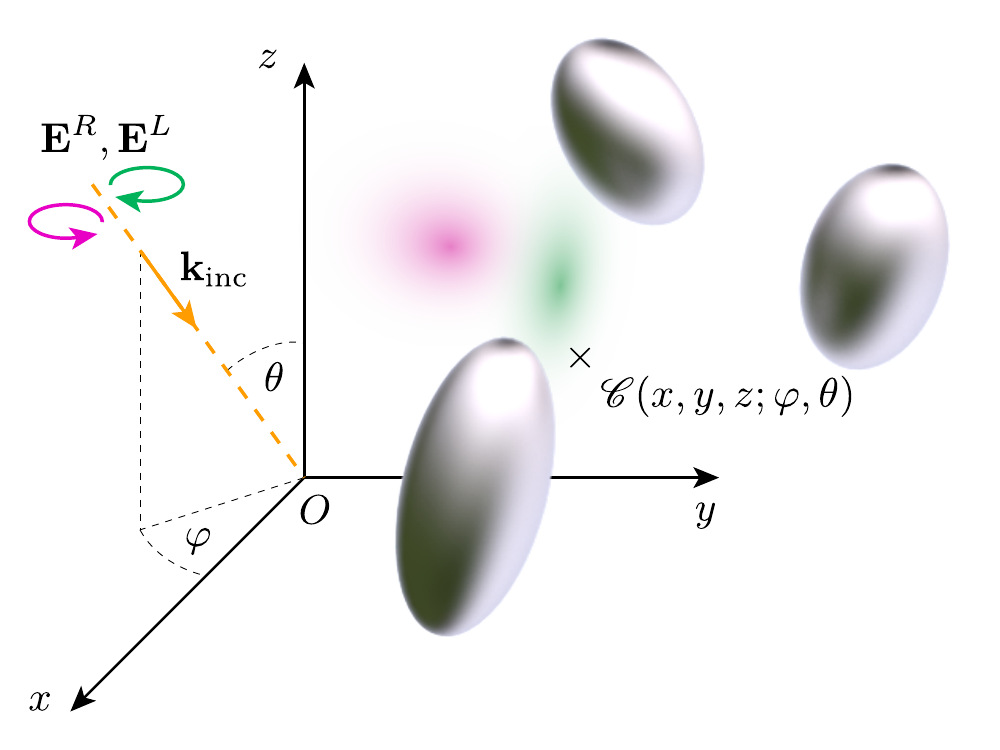}
	\caption{Schematic illustration of the system under study. A cluster of $N$ particles which are placed at arbitrary  positions and orientations, is illuminated by incident light in the form of a plane wave with wave vector $\mathbf{k}_\text{inc}$ indexed by Euler angles $(\theta,\varphi)$ with respect to a frame attached to the particle cluster. In practice, the particles would be held rigidly in place by a physical support (template); the support's potential influence on optical properties is assumed to be negligible in this work. We are interested in the variation of the cluster's optical properties when the incident light changes orientation. We compute far-field cross-sections (total extinction, scattering and absorption) and in particular the differential cross-sections associated with left and right-handed circularly-polarised light (circular dichroism). Another quantity of interest is the near-field local degree of optical chirality $\ldoc(x,y,z)$ for either left- or right-handed incident light.}
	\label{fig0}
\end{figure}

The theoretical description of many experiments requires averaging the optical response over all possible particle cluster orientations, as they randomly move in solution\cite{mishchenko2002scattering, Avalos-Ovando2021vv, xu2003orientation}. This averaging is often done numerically by simulating the optical response for several directions of incidence. For very small clusters, compared to the wavelength, it has been proposed that 3 directions of incidence (along 3 orthogonal axes, x, y, z) provide a good estimate of the orientation-averaged circular dichroism response\cite{Fan:2012tc,fan2013optical}. This approximation is expected to fail for larger clusters, but its range of validity has not been explored, leaving researchers to use their own numerical integration prescription in each particular study. Accurate orientation-averaging is particularly important for quantities relating to optical activity, i.e the differential response to left and right circularly polarised light, such as far-field circular dichroism, and near-field local degree of optical chirality. Conversely, the chiroptical response along specific indicence directions can differ markedly from the orientation-averaged one. This can have important implications when the clusters assume a preferential orientation, such as in many dark-field spectroscopy experiments with particles immobilised on a substrate and illuminated over a restricted solid angle.

We illustrate here the dependency of chiroptical properties on cluster orientation, and the associated difficulty in obtaining an accurate orientation-averaged response. Figure \ref{fig0} depicts schematically the light scattering problem under consideration, with a rigid cluster of $N$ particles (held together by a template, which is however neglected in the simulations) immersed in an infinite homogeneous medium (water or air), illuminated by a plane wave with left or right handed circular polarisation. The incidence direction is indexed by Euler angles $(\theta,\varphi)$ in a cartesian frame attached to the cluster. Our simulations are performed using a rigorous solution of Maxwell's equations for multi-particle clusters, the superposition \tmatrix\ method\cite{mishchenko2002scattering, StoutAL02, Schebarchov:2019aa, fazel2021orientation}. In this framework, incident and scattered fields are expressed in bases of vector spherical waves; the scattering properties of the cluster are thereby captured in a way that is independent of the incident field, leading to efficient calculations for multiple independent incident fields, and to analytical formulas for orientation-averaged properties. These provide accurate benchmark results against which the performance of numerical methods can be tested\cite{mishchenko2002scattering, mackowski1994calculation}.



%
\section{``Fingers crossed'' dimer}
%

\begin{figure*}[!htpb]
	\centering
\includegraphics[width=\textwidth]{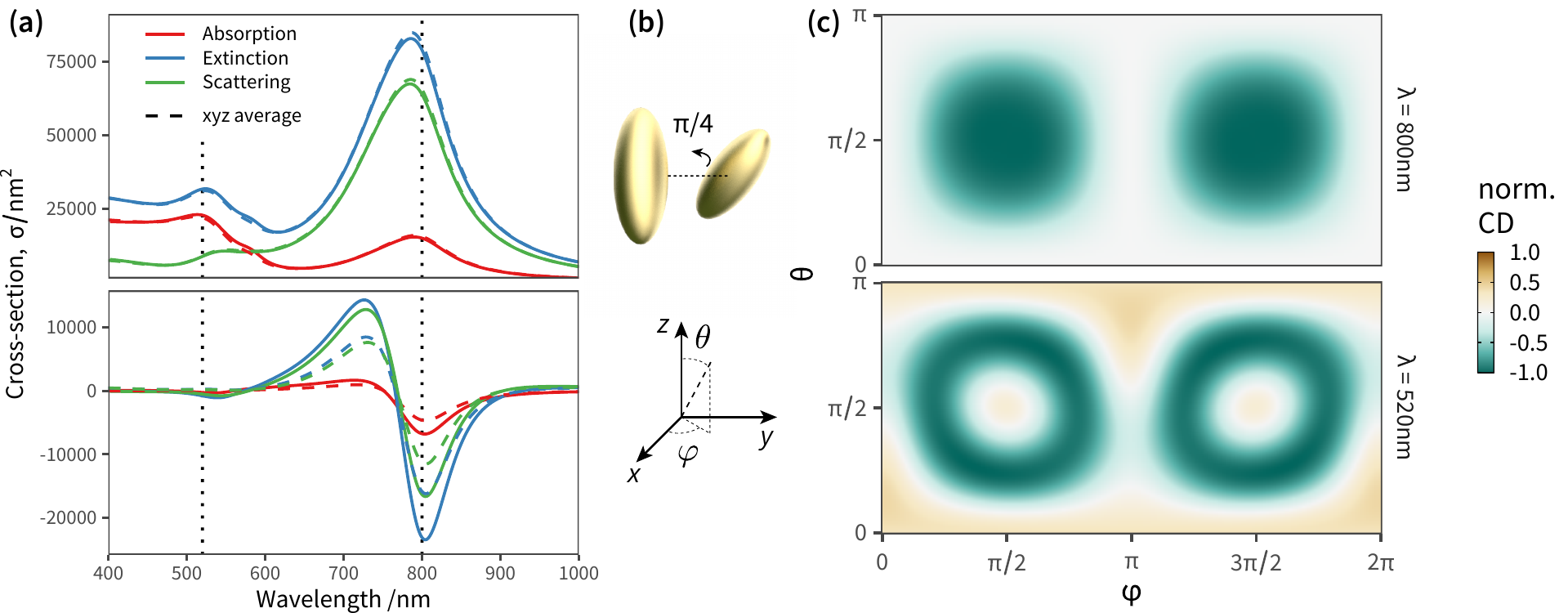}
	\caption{Angular response of a ``fingers crossed" dimer of gold spheroids immersed in water (refractive index 1.33). (a) Orientation-averaged far-field extinction (blue), absorption (red), and scattering (green) cross-sections. Lower panel: Orientation averaged circular dichroism spectra. The solid lines correspond to analytical orientation-avering (exact), while the dashed lines are for the numerical approximation using $x,y,z$ incidence directions only. (b) Schematic representation of the dimer of gold nanorods. Each particle is modelled as a prolate spheroid with semi-axes $a = b = 30$\,nm and $c = 80$\,nm. The first particle is at $(0,-100,0)$ with major axis along $z$, while the second particle is at $(0,+100,0)$, and titled from the $z$ axis by a polar angle $\theta = \pi/4$. (c) Angular map of the extinction CD at $\lambda = 800$\,nm (upper panel) and $\lambda = 520$\,nm (lower panel). The values are normalised to unity in both panels to facilitate comparison of the patterns. See ESI Fig.~S2$^\dag$ for maps at other wavelengths.}
	\label{fig1}
\end{figure*}

The first structure under consideration is a ``fingers crossed'' dimer of gold nanorods\cite{auguie2011fingers, zhao2017chirality} (Fig.~\ref{fig1}). This is arguably the simplest chiral nanostructure composed of achiral particles: with spherical particles, a minimum of 4 spheres is required\cite{mastroianni2009pyramidal, fan2013optical}. Many studies have explored the strong chiroptical response displayed by such dimers\cite{Dios:2019ug,Kuntman:2018vg,Wang:2015vi,Qu:2018wp}, with a characteristic bisignate lineshape around the dominant plasmon resonance of the nanorods\cite{auguie2011fingers, yin2013interpreting, najafabadi2017analytical, lan2013bifacial}. This signature circular dichroism is reminiscent of exciton-coupling between chromophores\cite{harada1978circular}, and is well-described in the coupled-dipole approximation. 

We can artificially scale the structure -- both the size of the particles, and their separation -- from a deep-subwavelength regime, dominated by absorption, to a scattering dominated lineshape when approaching the wavelength, and observe the evolution of the circular dichroism response (additional simulations in ESI Fig. S1$^\dag$). The plasmon resonances red-shift with increased size due to radiative damping, the relative contribution of scattering over scattering starts to dominate, but the lineshape remains very similar. As proposed by Govorov \emph{et al}, averaging the circular dichroism spectra over 3 orthogonal directions of incidence $x,y,z$ suffices to obtain a good accuracy compared to the exact orientation-averaged spectrum, for clusters that are small compared to the wavelength \cite{fan2013optical}. However, with larger particles and/or separation, light is able to excite non-negligible multipolar contributions to the far-field response of the cluster, which gradually increase the complexity of the angular excitation pattern particularly in the high-energy region, with a detrimental effect on the accuracy of a simple "$x,y,z$" angular averaging (Fig.~S1$^\dag$, $f>0.8$; Fig.~\ref{fig1} corresponds to the largest scale, $f=2$).

We compare in Fig.~\ref{fig1}c the angular maps at two wavelengths, $520$\,nm and $800$\,nm, corresponding to the vertical lines in the spectra of Fig.~\ref{fig1}a. These maps correspond to extinction circular dichroism at fixed incidence, with the wave-vector spanning the full solid angle. For visualisation purposes the values are scaled to unity, helping the comparison between the two wavelengths.  The circular dichroism is strongest in the spectral region of the main longitudinal plasmon resonance ($800$\,nm), and when light is incident along the dimer axis. Interestingly, circular dichroism vanishes when the wave vector is orthogonal to the dimer axis. The map at $520$\,nm displays a richer pattern, due to the contribution of higher-order multipoles, beyond the dipolar term, needed to describe the light scattered by the dimer. This leads, in turn, to a less accurate orientation average when using three directions only. 

\begin{figure}[!htpb]
	\centering
\includegraphics[width=\columnwidth]{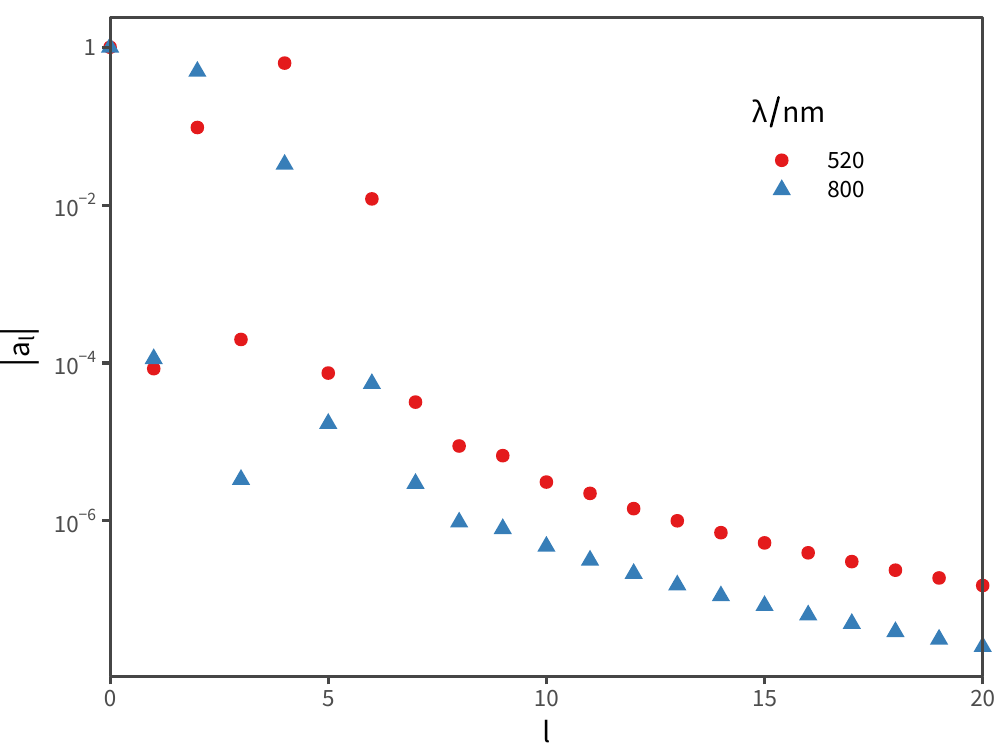}
	\caption{Spherical harmonic analysis of the circular dichroism angular patterns shown in Fig.~\ref{fig1}(c). We use the function \texttt{sphharm} from the Matlab library Chebfun\cite{T.-A.:2014wj,chebfun:2019aa} to compute the decomposition of the angular map into spherical harmonics. For each harmonic degree $l$ we collapse the corresponding $2l+1$ orders as  $|a_l|=\sum_{m=-l}^l |a_{lm}|^2$; the result summarises the total weight of degree $l$ in the pattern (higher degrees correspond to higher angular frequencies). For ease of comparison the scale is logarithmic, and we normalise the results such as $|a_0|=1$. The two cases considered correspond to $\lambda = 520$\,nm (red dots) and $\lambda = 800$\,nm (blue triangles).}  
	\label{fig2}
\end{figure}

To better quantify and contrast the relative complexity in the angular patterns of Fig.~\ref{fig1}c, we perform a spherical harmonic (SH) decomposition of the two-dimensional angular profile, using the Chebfun library\cite{T.-A.:2014wj}. Specifically, we project the far-field differential cross-section $\sigma_\text{\sc cd}(\theta,\varphi)$ onto spherical harmonics (SH) up to degree 20 as
\begin{equation}
\sigma_\text{\sc cd}(\theta,\varphi) \approx \sum_{l=0}^{20} \sum_{m=-l}^l a_{lm} Y_{lm}(\theta,\varphi),
\label{eq:sh}
\end{equation}
with $Y_{lm}$ the standard spherical harmonics of degree $l$ and angular momentum $m$\cite{mishchenko2002scattering}, and $a_{lm}$ the coefficients obtained by inner product of the angular pattern with the basis functions $Y_{lm}$\cite{chebfun:2019aa}. Roughly speaking, the SH decomposition performs a similar task as Fourier series for one-dimensional signals, but on the surface of a sphere. There is a close relationship, known as Jeans' formula, between the degree of spherical harmonics and the angular periodicity ("wavelength") of features appearing in the $(\theta,\varphi)$ pattern, allowing us to interpret the weight of higher-order spherical harmonics as the contribution of sharper angular features. We can also relate such weights to the difficulty in obtaining accurate angular averaging, as more directions of incidence will be required in the numerical cubature as the angular pattern presents higher-frequency features\cite{Penttila:2011ws}. The more complex pattern at $\lambda=520$\,nm compared to the main resonance at $800$\,nm is apparent in the relative weights of the first few coefficients, notably $l=2,4,6$. Based on the rapid fall-off of the coefficients, contributions up to $l\sim 6$ carry most of the information in the angular dependence, and we can predict a cubature rule with a dozen points to be very accurate\cite{lebedev1976quadratures}. Indeed, just the 3 orthogonal x, y, z directions provide an accurate orientation average for the smaller dimers tested (see ESI Fig. S1$^\dag$, panels $f<1$), and a correct spectral shape even for the larger dimer (ESI Fig. S1$^\dag$, panel $f=2$, also in the main text Fig.~\ref{fig2}(a)).


%
\begin{figure*}[!htpb]
	\centering
\includegraphics[width=\textwidth]{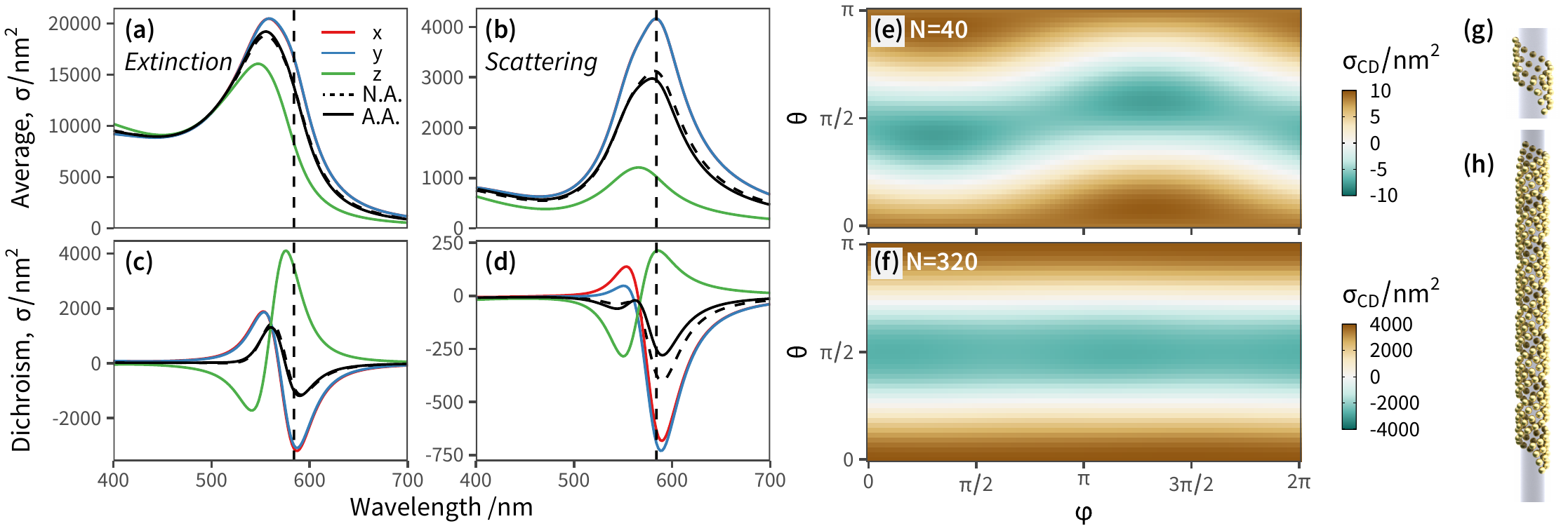}
	\caption{Far-field chiroptical response of a ``GoldHelix'' cluster described in Ref.~\textcite{cheng2017goldhelix} consisting of a ribbon of 5\,nm-radius Au nanospheres wrapped in a helix and immersed in water. The list of particle positions we used is available in ESI$^\dag$. (a, b) Extinction and scattering spectra for three perpendicular incident directions, $x$-axis (solid red line), $y$-axis (solid blue line), $z$-axis (solid green line), the average over the three (dashed black line), and the analytical orientation averaged result (solid black line). (c, d) The corresponding results for circular dichroism. (e, f) Angular map of circular dichroism (for extinction) for a GoldHelix with number of particles $N = 40$ (e) and $N = 320$ (f). (g, h) Schematic illustration of the GoldHelix geometry with $N = 40$ and $N = 320$, respectively. Note that cross-sections are normalised by the number of particles in each cluster, to facilitate comparisons. The simulations considered a maximum multipole order $l_\text{max}=4$.}
	\label{fig3}
\end{figure*}
\section{Helix of nanoparticles}

The dimer example above is a very compact cluster, with only two particles. With more particles, the optical response becomes even richer, as the resonances hybridise in complex ways\cite{Hentschel:2011vj,lan2015nanorod,hentschel2012three}. This is particularly striking in the circular dichroism response of small helices of 5--10 nanospheres, where the addition of one particle can flip entirely the CD spectrum\cite{guerrero2011individual,Fan:2011ty}. This sensitivity to precise positioning can potentially be an asset, enabling very precise retrieval of geometrical information from far-field optical interrogation, or a drawback if the fabrication technique does not offer sufficient control to ensure precisely identical clusters. We note that helices with different parameters, or with more particles, can be more robust\cite{Fan:2011ty}. Alternatively, replacing the spherical particles with elongated nanorods aligned with the helix, results in a robust CD lineshape\cite{guerrero2011individual}. This is because the induced electric dipole for each particle is given a preferential orientation, unlike with nanospheres where its orientation is purely dictated by the self-consistent field originating from the incident light and the field scattered by its neighbours. This serves to illustrate the complex interplay of relative phase and direction of the vector fields, at the heart of chiroptical responses in nanoparticle assemblies. Helices appear commonly in natural systems presenting optical activity, and are indeed a prototypical structure to study the interaction of light with chiral matter. Early theoretical models based on coupled-dipole approximations have considered the response of helices, and notably the possibility to induce long-range interactions through the well-defined phase relationship between induced dipoles positioned in such geometries\cite{Applequist:1979tp}.

In the more recent context of chiral clusters of plasmonic particles, long helices have also been proposed\cite{song2013tailorable,jung2014chiral,Lan:2018wq}, notably ``GoldHelix''\cite{cheng2017goldhelix} which comprise hundreds of gold nanospheres attached on a helical silica template. This structure provides an interesting example of a relatively large rigid cluster of nanoparticles, which is still free to move and rotate randomly when characterised, or used, in solution. We simulated a representative ``GoldHelix'' cluster of Au nanospheres 5\,nm in radius, following the description in Ref.~\textcite{cheng2017goldhelix}, with $4\times 80$ nanospheres arranged in a $\sim 470$\,nm-long helix (Fig.~\ref{fig3}; see ESI$^\dag$ for the list of particle positions we used). Coupled-dipole modelling described in \textcite{cheng2017goldhelix} indicated the need for 30 directions of incidence over an octant to accurately model the averaged circular dichroism spectra. We explore the reasons for this relatively challenging numerical averaging by simulating the differential CD response of the helix as a function of $\theta$ and $\varphi$ angles. Note that we chose to orient the axis of the helix along $z$, for easier interpretation (Fig.~\ref{fig3}g,h).

To fully illustrate the influence of the cluster's spatial extent and anisotropy, we compare two structures: the full helix (Fig.~\ref{fig3}f,h), and a truncated helix with identical parameters but only $4\times 10$ particles (Fig.~\ref{fig3}e,g). The optical response is qualitatively similar with a strong bisignate CD spectrum at the position of the localised plasmon resonance ($\lambda\approx 550$\,nm) of small Au nanospheres in water. The extinction spectrum is dominated by absorption, but we note that some characterisation techniques such as dark-field spectroscopy would measure a somewhat distinct lineshape (panel d). Both structures present an angular dependence with a strong pattern in the polar angle $\theta$, but only a small variation along $\varphi$ which becomes almost invisible for the longer helix. 
Circular dichroism is strongest for incidence with $k$ vector along the helix axis, as would be expected since both the helix of the structure and of the circularly polarised light overlap. Perhaps less intuitively, incidence normal to the helix axis also yields intense circular dichroism, and reverses its sign. Interestingly, the circular dichroism vanishes at $\theta=\pm \pi/4$ for this helix; at such angles, the "ribbon" of nanospheres upon which the (idealised) structure is built appears locally as a square array, and the near-neighbour electromagnetic couplings are therefore achiral. 

Because GoldHelix clusters are quite long ($\sim 470$\,nm in this example) and have only few particles breaking the rotational symmetry at both ends, the variation with azimuth is much less pronounced, and the differential angular response is relatively independent of $\varphi$. We can perform the same spherical harmonic decomposition as described above for the dimer, with the results shown in ESI Fig.~S4$^\dag$. The spherical harmonic coefficients decay less rapidly, and the larger GoldHelix would typically require a degree $l\approx 10$ to capture most of the angular pattern. In terms of numerical orientation-averaging, optimal cubature methods such as Lebedev cubature would require 40 to 50 points to integrate exactly spherical harmonics of degree 10\cite{Penttila:2011ws}, justifying the relative difficulty in estimating numerically the orientation average for the coupled-dipole simulations in Ref.~\textcite{cheng2017goldhelix}. We note that the \tmatrix\ framework offers clear advantages for this type of situation, where access to the analytical result for orientation-averaged optical properties provides a rigorous benchmark to assess the accuracy of numerical cubature methods. A detailed comparison of the performance of different cubature methods for light scattering calculations is however beyond the scope of this work, and will be presented elsewhere.


%
\section{Local degree of chirality: Si nanospheres}
\begin{figure*}[!htpb]
	\centering
\includegraphics[width=\textwidth]{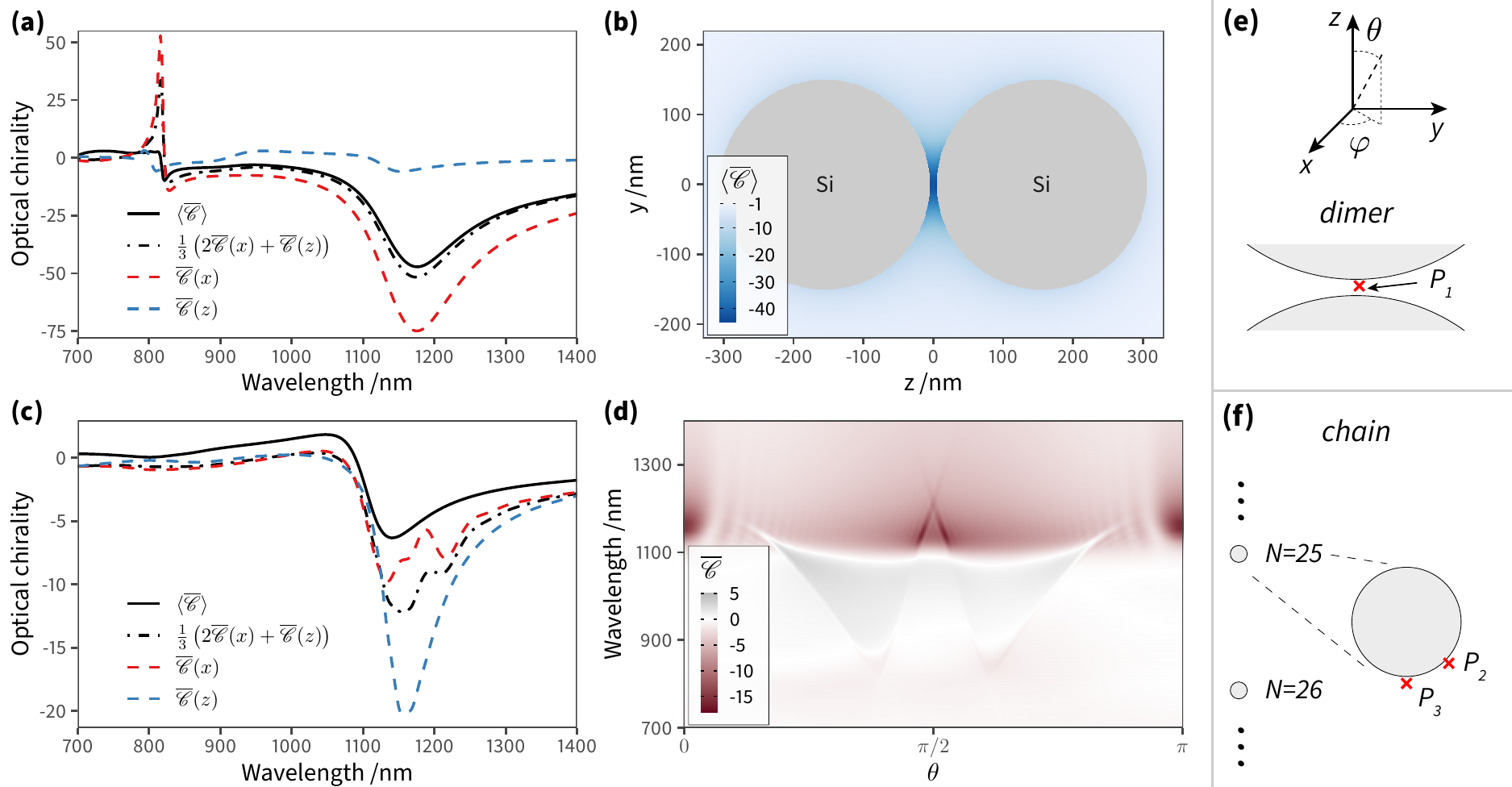}
	\caption{Normalised degree of optical chirality around Si nanospheres of radius $r = 150$\,nm, for a dimer with inter-particle $\text{gap} = 10$\,nm and a diffractive chain of 50 Si nanospheres with inter-particle distance $d = 1208$\,nm. The particles are placed in air. (a) Spectrum of $\ldocbar$ at the point $P_1$ in the gap between two spheres, $4$\,nm away of the sphere's surface, as depicted in schematic (e). The structure is illuminated with a right-handed circularly-polarised plane wave incident along the $x\equiv y$ direction (dashed red line), or $z$ direction (dashed blue line). The numerical average over these perpendicular incident directions (N.A., dash-dotted black line) is compared with the analytical orientation-averaged spectrum $\ldocoabar$ (A.A., solid black line). (b) Map of $\ldocoabar$ around the dimer, calculated in the $y-z$ plane at the wavelength $\lambda = 1175$\,nm, corresponding to the strongest value of $\ldocoabar$ in the spectrum. (c) Spectrum of $\ldocbar$ at the off-axis point $P_2$ near particle 25 in the 50-particle chain, as depicted in (f). The illumination directions are the same as (a). (d) Dispersion map of $\ldocbar$ at the point $P_3$, varying the incident wavelength and direction $\theta$.} 
	\label{fig4}
\end{figure*}

Our last example moves away from plasmonic nanoparticles and considers instead silicon dielectric spheres. With a relatively high refractive index\cite{palik1998handbook}, silicon nanoparticles have recently attracted attention for their ability to sustain intense electric and magnetic resonances in the visible--near-IR region\cite{evlyukhin2012demonstration, evlyukhin2011multipole}. Of particular interest in the context of chiroptical properties is the recent realisation that such scatterers can conserve light's helicity\cite{lasa2020surface} in the vicinity of the Kerker condition (electric and magnetic dipole coefficients of equal magnitude). This property leads to locally-enhanced near-fields at resonance without change in handedness, that would otherwise compromise or lessen the amplification of signals for chiroptical spectroscopies of molecular adsorbates\cite{Poulikakos:2019ts}. We focus here on a near-field quantity, namely the local degree of optical chirality $\ldoc\propto\Im(\mathbf{E}^*\cdot\mathbf{B})$, with $\mathbf{E},\mathbf{B}$ the complex electric and magnetic fields. In the plots below we display $\ldocbar=2\ldoc/(k\varepsilon_0E_0^2) $, normalised with respect to the value of circularly-polarised plane waves with incident field $E_0$. The quantity $\ldocoabar$ denotes the orientation-average of $\ldocbar$ with respect to the direction of incident light\cite{fazel2021orientation}.
%

We assess the angular dependence of the local degree of optical chirality at small distances from a dimer of Si nanospheres (radius $r = 150$\,nm and $\text{gap} = 10$\,nm), where multipolar modes can play an important role in the optical spectrum. Figure~\ref{fig4}(a) presents the normalised optical chirality $\ldocbar$ at position $P_1$ in the gap. The results are shown for two perpendicular incident directions, $k_x\equiv k_y$ and $k_z$, from which we compute the numerical average over three perpendicular incident directions (dash-dotted black line), to contrast with the analytical orientation-averaged result $\ldocoabar$ (solid black line). These calculations used a maximum multipolar order $l_\text{max}=12$ in the \tmatrix\ method, sufficient to capture the role of higher order modes. 
Although the average over three perpendicular incident directions is relatively similar to $\ldocoabar$ at lower energies, where the dimer's response is dominated by dipolar electric and magnetic resonances, at higher energies there is a noticeable difference. In fact, the sign of the averaged $\ldocbar$ and $\ldocoabar$ is different at the shorter wavelengths. The interaction between quadrupoles and other higher order modes in the Si nanospheres results in a distinct sharp peak with a Fano resonance lineshape at $\lambda = 816$\,nm, excited when the incident wave vector is perpendicular to the main axis of the dimer ($k_x\equiv k_y$). Although it is often the case that dielectric nanoparticles preserve the sign of the helicity of the incident wave, this is not observed here in this spectral region.

It is the case, however, for a dimer or indeed any arrangement of such scatterers, that the helicity of incident light is preserved in the scattering process\cite{lasa2020surface, hanifeh2020helicity} near the Kerker condition\cite{kerker1983electromagnetic, nieto2011angle, zambrana2013duality}. This is apparent in the map of $\ldocoabar$ at $\lambda = 1175$\,nm, where only negative values appear (Fig.~\ref{fig4}b; the Kerker condition is at $\lambda = 1208$\,nm, see ESI Fig.~S4$^\dag$). Molecules adsorbed anywhere on the surface of this particle cluster will experience an enhanced degree of local optical activity, compared to the value $(-1)$ corresponding to the excitation in free-space by a right circularly-polarised plane wave. In contrast, a similar dimer with plasmonic particles would typically show a spatial distribution of peaks of positive and negative signs, weakening the spatial average for analytes uniformly adsorbed on the particle surfaces\cite{fazel2021orientation}. Remarkably, $\ldocoabar$ is enhanced by a factor $\sim 44$ in the gap between nanospheres, compared to the incident plane wave. This enhancement reaches $\sim 75$ times for the fixed incident direction along $k_x\equiv k_y$. Si  dimers are therefore a good candidate for producing superchiral fields of use in surface-enhanced spectroscopy of chiral molecules\cite{TangCohen2010, lasa2020surface}.

The Si dimer, much like the Au dimer presented above, presents a relatively smooth angular dependence, due to its compactness. There is however a richer spectral signature due to the excitation of higher-order resonances. 
For our last example, we consider a recently-proposed variation on the dimer of dielectric resonators, in the form of a long chain of Si nanospheres (Fig.~\ref{fig4}(c) with $N=50$, in our case, but longer chains have been studied). This provides a realistic example of an extended, one-dimensional structure\cite{lasa2020surface}, bridging the regimes of compact, solution-based clusters, and of more extended structures obtained by top-down lithography. Periodic chains (or arrays) of resonant particles have received a lot of attention in the context of plasmonics and meta-surfaces, notably for their ability to support intense resonances when the periodicity of the chain is commensurate with the light wavelength. This occurs with the coherent superposition of scattering events at each particle along the chain, i.e. the condition for a grazing diffraction order. In this regime, the particles can be strongly coupled over very large distances despite their wide separation, and sustain a highly-delocalised mode. The inter-particle distance is here chosen to coincide with the Kerker condition ($\lambda = 1208$\,nm), to combine the benefits of field enhancement and helicity preservation\cite{lasa2020surface}.

The elongated GoldHelix structure already illustrated that as a cluster structure extends in an anisotropic way, along at least one direction, the collective optical response becomes more dependent on the incidence direction. This can be understood from the perspective of a multipolar expansion, as used in the superposition \tmatrix\ framework\cite{mackowski1994calculation,StoutAL02}. Placing the origin of coordinates somewhere near the centre of mass of the particle cluster, the description of the scattered field from the cluster as a whole requires higher multipole orders as the size parameter of the cluster increases, that is, the wavenumber times the radius of the cluster's circumscribed sphere.
In turn, higher multipole orders are associated with higher frequencies in the angular pattern of the scattered field, and, by optical reciprocity, of the incident field causing the local enhancement $\ldocbar$ or the far-field response.

Figure~\ref{fig4}(f) depicts the chain with 50 Si nanospheres along the $z$ axis, with separation $d = 1208$\,nm; the degree of optical chirality $\ldocbar$ and $\ldocoabar$ is calculated for an off-axis point $P_2$ placed at $\theta = \pi/3$ and $ 5$\,nm away of the middle particle's surface (Fig.~\ref{fig4}(c)). The Si chain yields a maximum value of optical chirality $\ldoc$ at $\lambda = 1156$\,nm and $\lambda = 1134$\,nm for incident directions along $k_z$ and $k_x\equiv k_y$. We note that for this chain, the peak of the $\ldoc$ spectrum is mainly a result of the interaction between electric and magnetic dipoles induced in the particles, with negligible influence of higher order modes. Remarkably, there is a clear difference between the maximum value of $\ldocoabar$ and the average over three perpendicular incident directions. One might have expected that the orientation-averaged spectrum should fall somewhere between the three perpendicular incident directions ($x,y,z$), yet here $\ldocoabar$ is in a different range altogether. To better understand this counter-intuitive result, we present in Fig.~\ref{fig4}(d) a 2D map of $\ldocbar$ versus wavelength and polar angle $\theta$, i.e. a dispersion plot. Note that for this map we chose a position $P_3$ along the $z$ axis so that $\ldocbar$ is independent of $\varphi$, for simplicity. Varying the incident direction $\theta$ near the normal $\theta\sim\pi/2$, the diffractive mode moves very rapidly with angle and the strong value of $\ldocbar$ associated with it is not well captured in the three directions $x, y, z$.      
\section{Conclusions}
%

Although the angular dispersion of electromagnetic resonances is ubiquitous in the field of metamaterials, metasurfaces and photonics in general, it is much less often discussed or studied in compact clusters of nanoparticles, such as those self-assembled by chemical methods. Such clusters may be characterised in solution -- in which case the optical properties are typically averaged over all possible orientations over the duration of a measurement --, or probed over a finite solid angle of illumination, if immobilised on a subtrate, as in many microscopy techniques such as dark-field spectroscopy. For very small clusters, the common assumption that the orientation-averaged response can be obtained by averaging the response along x, y, z axes is generally valid, though the average may still hide some dramatic variation between the three cases. This is particularly clear for chiroptical properties such as circular dichroism, where the typically minute difference in scattering or absorption between left- and right-polarisations is very sensitive to the exact phase and polarisation relationships between the neighbouring particles and the incident field. As the cluster size grows, in comparison to the wavelength, a larger number of multipolar components is required to describe the optical response of the ensemble, seen as a single composite scatterer; in turn, these higher order components are associated with spherical harmonics of increasing degree and therefore angular frequencies. This leads to more complex angular patterns, as we illustrated for the case of a wavelength-sized helical structure. Finally, as the structure extends even further, the angular response can show sharply peaked angular features associated with electromagnetic modes of well-defined momentum. We illustrate this effect with a long chain of Si nanoparticles supporting a diffractive mode, where the excitation of the mode at specific phase-matching conditions leads to an enhanced optical response, here the local degree of optical chirality. 

This brief selection of examples should illustrate the importance of considering orientation dependence in the characterisation and modelling of clusters of nanoparticles, even those fabricated by bottom-up synthesis routes. Often, the numerical simulation methods can discourage the comprehensive study of this angular response, as the time-consuming calculations need to be repeated for many angles of incidence, but we hope our benchmark results on select configurations provide a useful guide for the study of similar geometries. Similarly, with experiments on immobilised clusters using a restricted angular range of illumination, the characterisation of chiroptical properties can easily become misleading due to extrinsic optical activity, i.e. a differential response to left and right polarisations induced by the specific direction of the incident wave vector. Unlike other optical properties such as field enhancement, scattering or absorption, with chiroptical properties the measured response in such situations may not only have a different magnitude, compared to the orientation-averaged response, it can even have the opposite sign altogether. 


%
\section*{Author Contributions}
All authors contributed to the design of the study, to the numerical simulations, and to the writing of the manuscript.

\section*{Conflicts of interest}
There are no conflicts to declare.

\section*{Acknowledgements}
The authors would like to thank Dmitri Schebarchov and Eric Le Ru for helpful discussions, the Royal Society Te Ap\=arangi for support through a Rutherford Discovery Fellowship (B.A.), and the MacDiarmid Institute for additional funding (A.F.N.).



\balance


\bibliography{references} 
\bibliographystyle{rsc} 

\end{document}


\maketitle
\tableofcontents

\section{Dimer of Au nanorods}

\begin{suppfigure}[!htpb]
\includegraphics[width=0.9\textwidth]{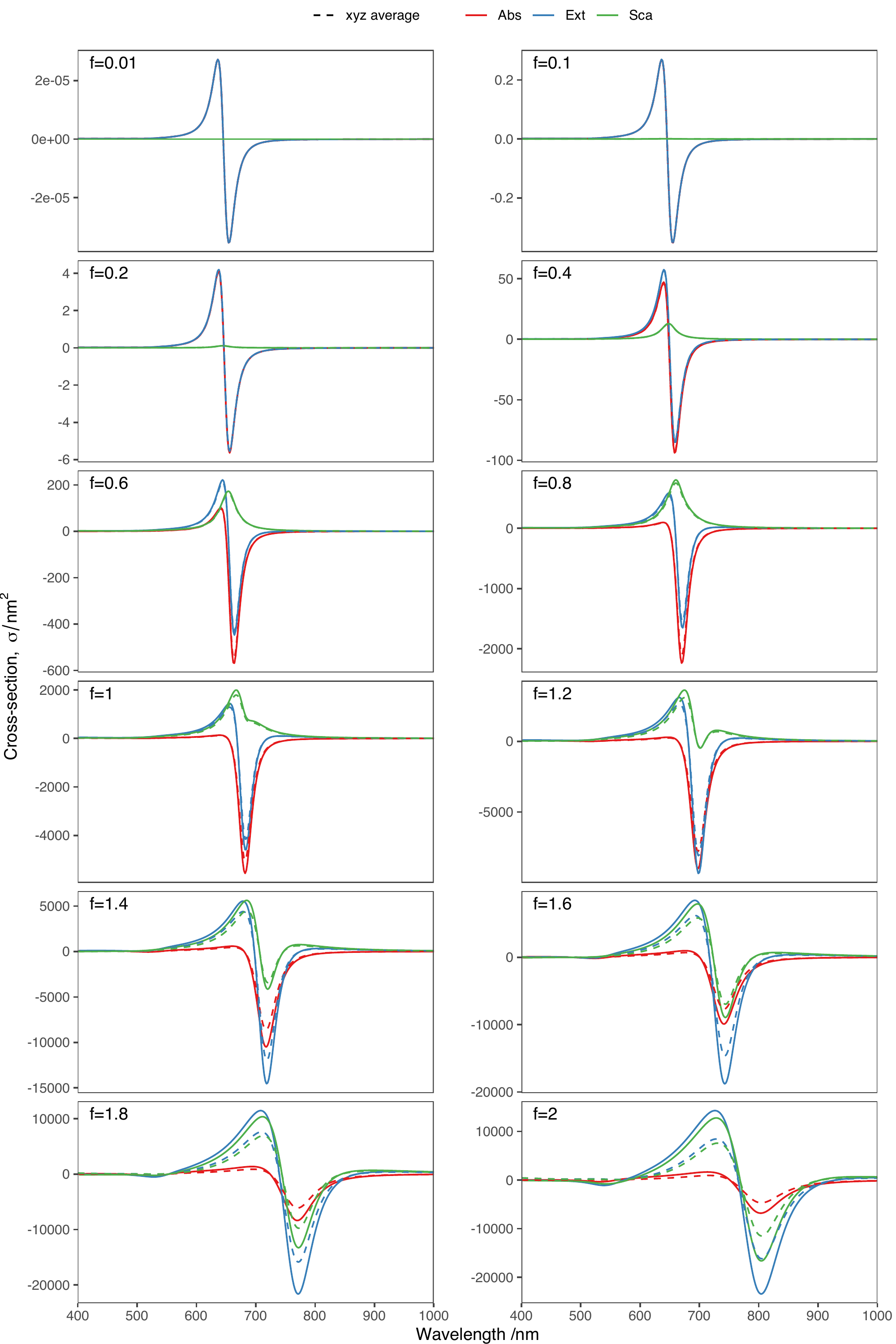}
    \caption{Absorption (red), scattering (green), and extinction (blue) circular dichroism spectra for the ``fingers crossed'' dimer of Fig.~2, with different scaling factor ($f$). The solid lines show the analytical orientation averaged results and the dashed lines are for the average over $x, y, z$. Each particle in the dimer is represented as a prolate ellipsoid with semi-axes $f.a$ and $f.c$ ($a = 15$\,nm and $c = 40$\,nm) and center-to-center separation $f.d$ ($d = 100$\,nm).}
	\label{figS1}
\end{suppfigure}

Figure~S\ref{figS1} presents the analytical orientation averaged results (solid lines) and the numerical averaging over $x,y,z$ incident directions (dashed lines) for absorption, scattering, and extinction circular dichroism spectra in the ``fingers crossed'' dimer configuration, with different scaling factor ($f$). The structure comprises two identical Au spheroids with semi-axes $f.a$, $f.b$, with nominal dimensions $a=15$\,nm, $c=40$\,nm, and centre-to-centre separation $f.d$, with $d=100$\,nm. For small scaling factors ($f\ll 1$), the dimer is much smaller than the wavelength and averaging over just the 3 orthogonal $x, y, z$ directions provides an accurate orientation averaging. However, the accuracy suffers as the scaling factor increases beyond $f \sim 0.8$. 

We also note that for small dimers, extinction is largely dominated by absorption and the circular dichroism spectrum has an antisymmetrical line shape, very reminiscent of exciton coupling in molecular CD. For larger scaling factors, scattering contributes more to the overall extinction CD with an increase in the intensity of the CD spectrum and a deviation from the antisymmetric curve predicted by exciton coupling theory. The intermediate regime where absorption is on par with scattering leads to interesting spectral profiles, such as a transition in the scattering lineshape from a monosignate band below $f=1.2$, evolving to a bisignate band resembling more that of absorption at larger sizes\cite{auguie2011fingers}.
 
 \clearpage
 
\begin{suppfigure}[!htpb]\centering
\includegraphics[width=\textwidth, keepaspectratio]{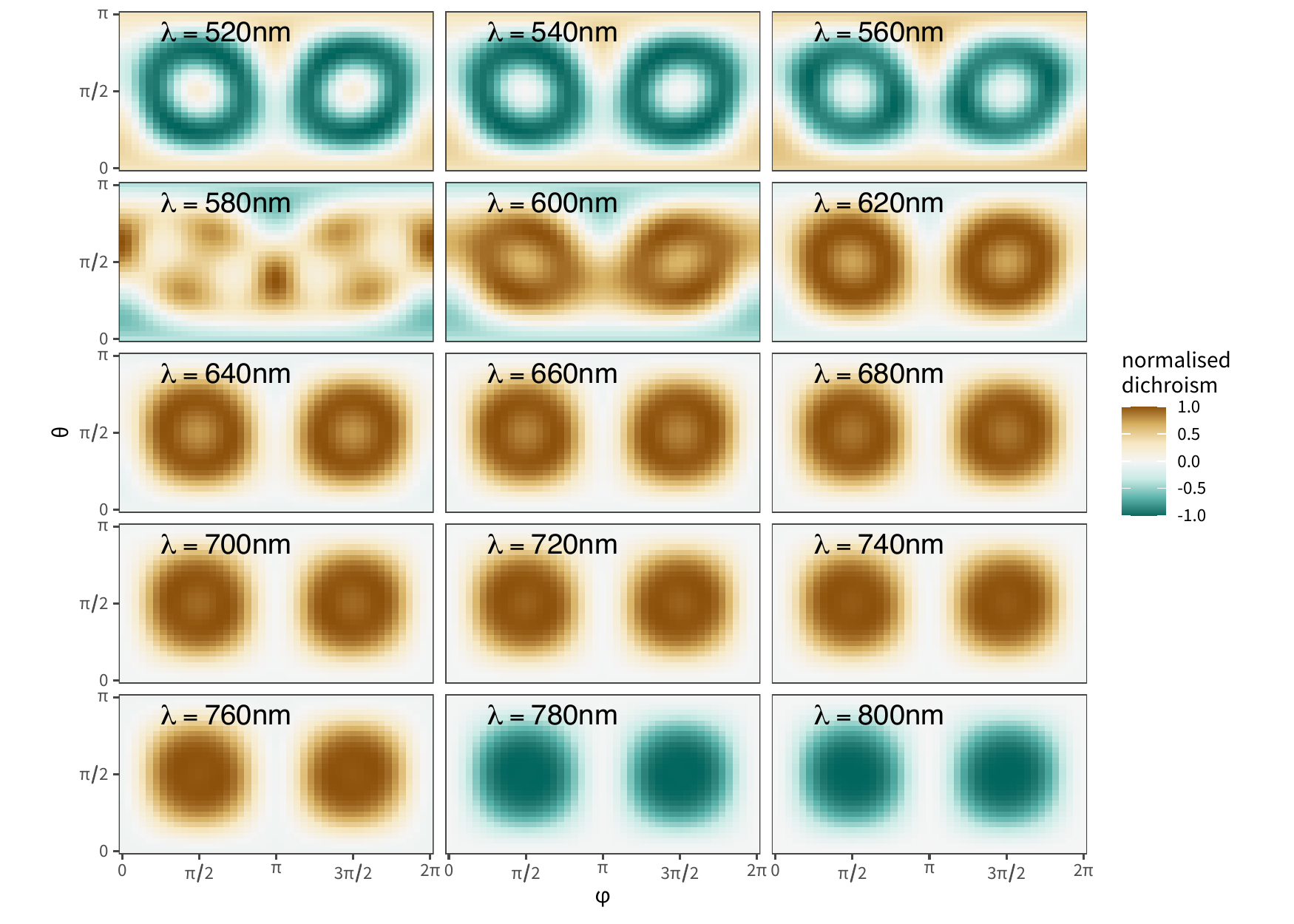}
    \caption{Angular maps of the extinction CD at different wavelengths for the ``fingers crossed" dimer consisting of two Au prolate spheroids with semi-axes $a = 30$\,nm and $b = 80$\,nm, interparticle separation $d = 200$\,nm (scaling factor $f=2$ in Fig.~S\ref{figS1}, corresponding to Fig.~2 of the main text), and relative orientation $\theta = \pi/4$.}
	\label{figS2}
\end{suppfigure}

Figure~S\ref{figS2} presents multiple maps of the angular pattern of extinction circular dichroism for the dimer ($f=2$) presented in the main text, Fig.~2. The different panels are for wavelengths spanning the spectral range 520--800\,nm, showing the evolution between the two extremes shown in the main text. Note that the colour scale is normalised to unity in each panel, to facilite comparisons of the patterns, as the absolute intensities vary strongly across the spectrum. 

\clearpage
\section{GoldHelix}
\begin{suppfigure}[!htpb]\centering
\includegraphics[]{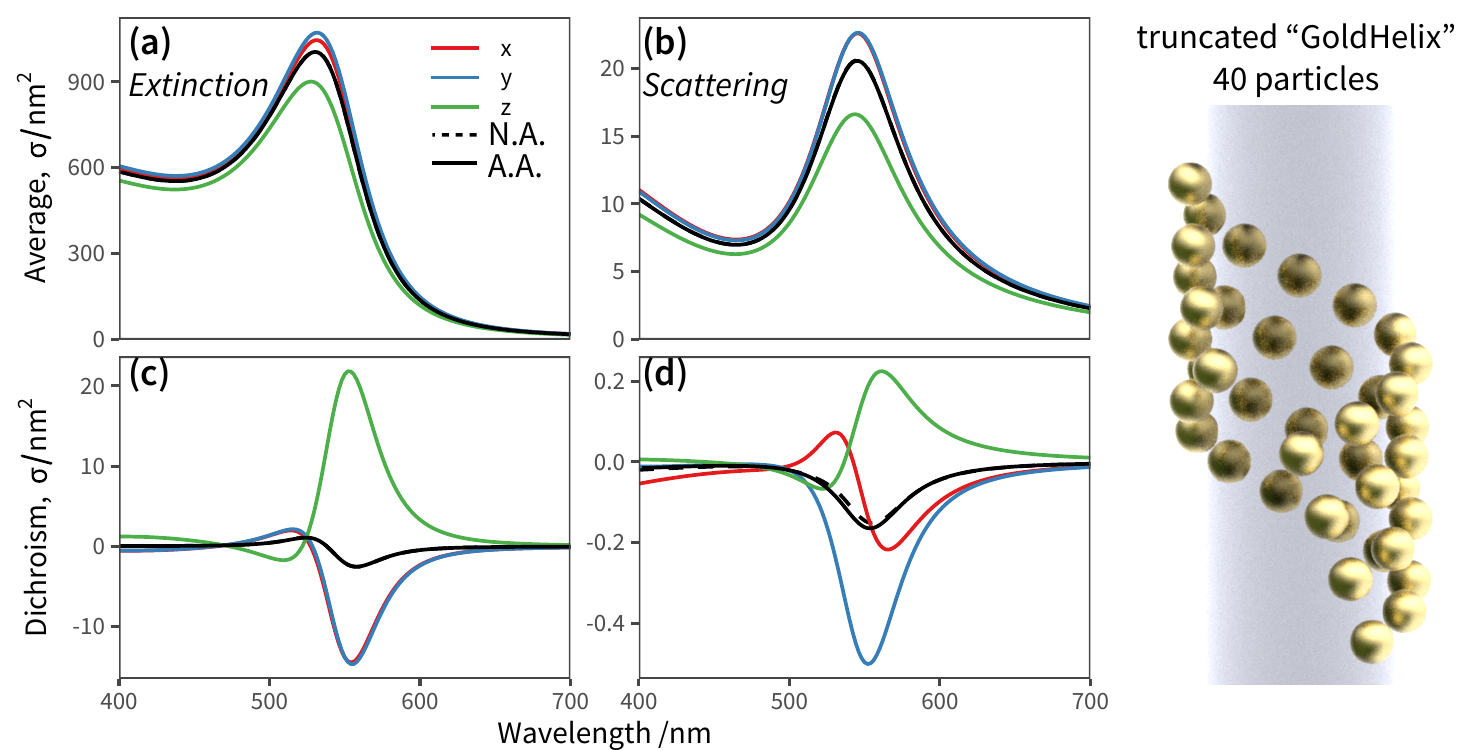}
    \caption{Angular response of a GoldHelix cluster\cite{cheng2017goldhelix} with $N =40$ particles ($4\times 10$ ribbon). Upper panels: Extinction and scattering spectra for three perpendicular incident directions, $x$-axis (solid red line), $y$-axis (solid blue line), $z$-axis (solid green line), the numerical average over the three (N.A., dashed black line), and the analytical orientation-averaged result (A.A., solid black line). Lower panels: the corresponding results for circular dichroism. Note that the numerical and analytical orientation averages are mostly indistinguishable, except for panel (d).}
	\label{figS3}
\end{suppfigure}

Figure~S\ref{figS3} complements Fig.~4 (a--d) of the main text, with far-field spectra for the truncated GoldHelix instead of the full structure. The 3D view on the right illustrates the geometry in more detail: a $4\times 10$ ribbon is formed by a square array of Au nanospheres, wrapped in a helix around a central template\cite{cheng2017goldhelix}. The template is not part of the electromagnetic simulations, however its effect on optical properties is expected to be relatively weak, as it is made of silica.

\begin{suppfigure}[!htpb]\centering
\includegraphics[]{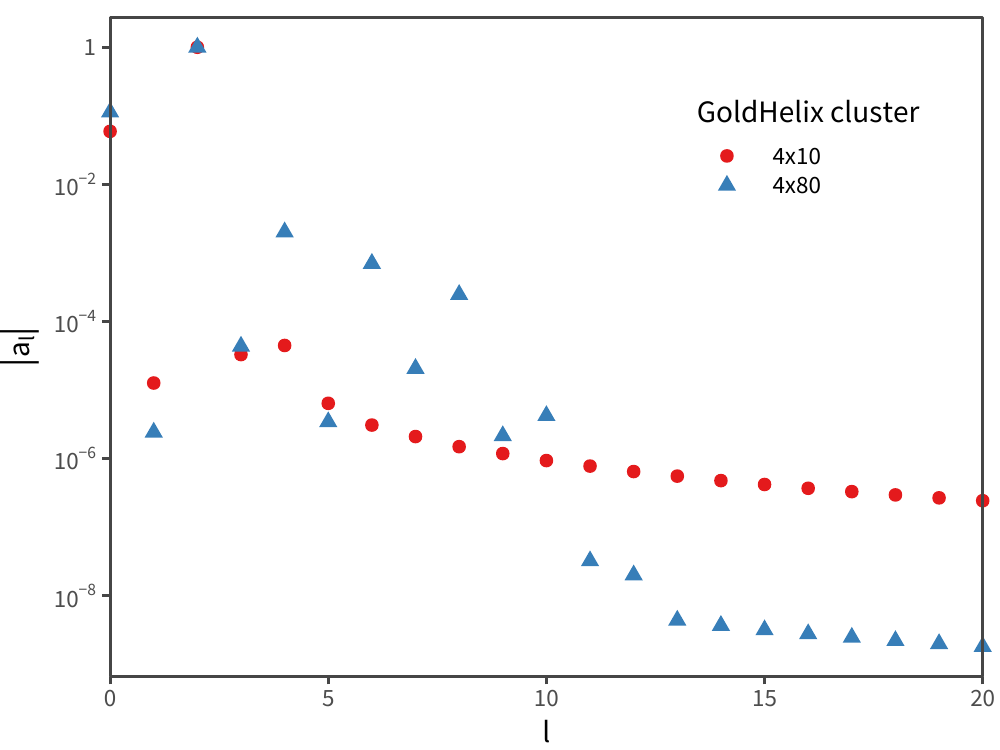}
    \caption{Spherical harmonic analysis of the circular dichroism angular patterns shown in Fig.~4 (e,f). We use the function \texttt{sphharm} from the Matlab library \texttt{chebfun}\cite{T.-A.:2014wj,chebfun:2019aa} to compute the decomposition of the angular map into spherical harmonics. For each harmonic degree $l$ we collapse the corresponding $2l+1$ orders as  $|a_l|=\sum_{m=-l}^l |a_{lm}|^2$; the result summarises the total weight of degree $l$ in the pattern (higher degrees correspond to higher angular frequencies). For ease of comparison the scale is logarithmic, and we normalise the results such as $\text{max}a_l=1$. The two cases considered correspond to the full GoldHelix cluster (blue triangles, 320 particles), and the truncated structure (red circles, 80 particles).}
	\label{figS4}
\end{suppfigure}

\clearpage
\section{Si chain}
Figure~S\ref{figS5} shows the extinction cross section of a single Si nanosphere with radius $r = 150~\text{nm}$ (black solid line). The spectra ED (red dashed line) and MD (blue dashed line) refer to electric dipole and magnetic dipole contributions to the extinction cross section, separately. It can be seen that spectra of ED and MD intersect at $\lambda = 1208~\text{nm}$ corresponding to the so called first Kerker condition.

Fig.~S\ref{figS6} shows the normalised extinction efficiency $Q_\text{ext} = C_\text{ext} / (N.C_\text{ext,1})$ for a chain of $N = 50$ Si nanospheres, where $C_\text{ext,1}$ is the extinction cross section of a single silicon particle at $\lambda_\text{Kerker}$. The structure is illuminated normal to the chain axis with a circularly polarised plane wave at $\lambda=1208$\,nm, and we vary the interparticle distance (pitch).  
\begin{suppfigure}[!htpb]
	\centering
\includegraphics{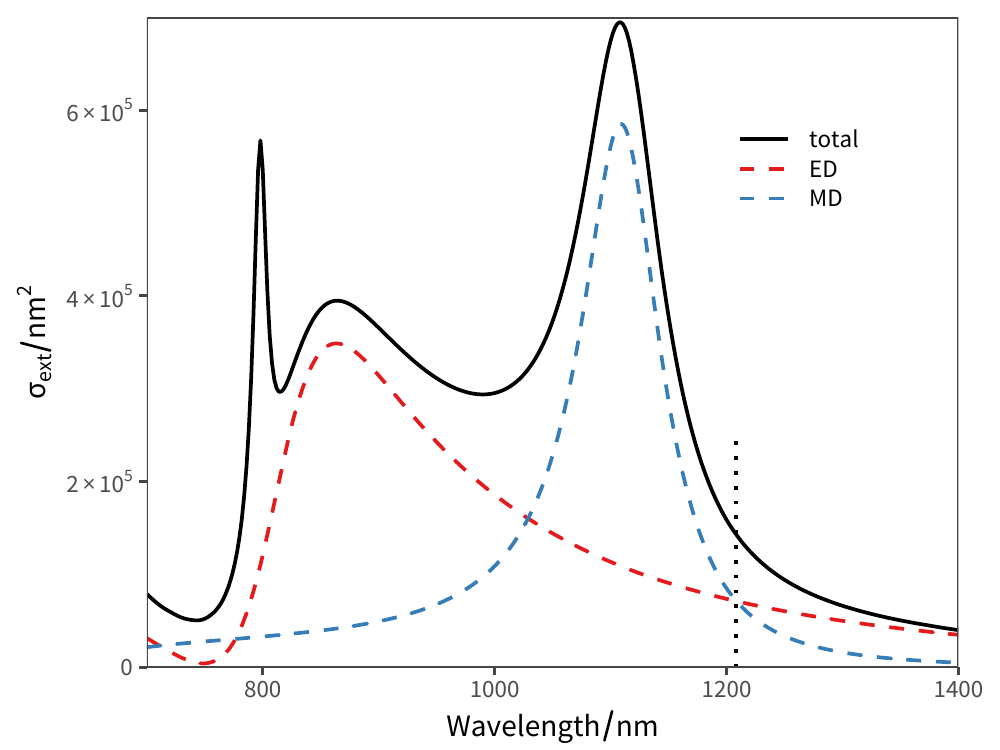}
    \caption{ Extinction spectrum of a single Si nanosphere with radius $r = 150$\,nm in air. The total spectrum (solid black line) is decomposed into electric dipole (ED, dashed red line) and magnetic dipole (MD, dashed blue line). A vertical dotted line at $\lambda=1208$\,nm indicates the Kerker condition where electric dipole and magnetic dipole terms are equal.}
	\label{figS5}
\end{suppfigure}  
\begin{suppfigure}[!htpb]
	\centering
\includegraphics{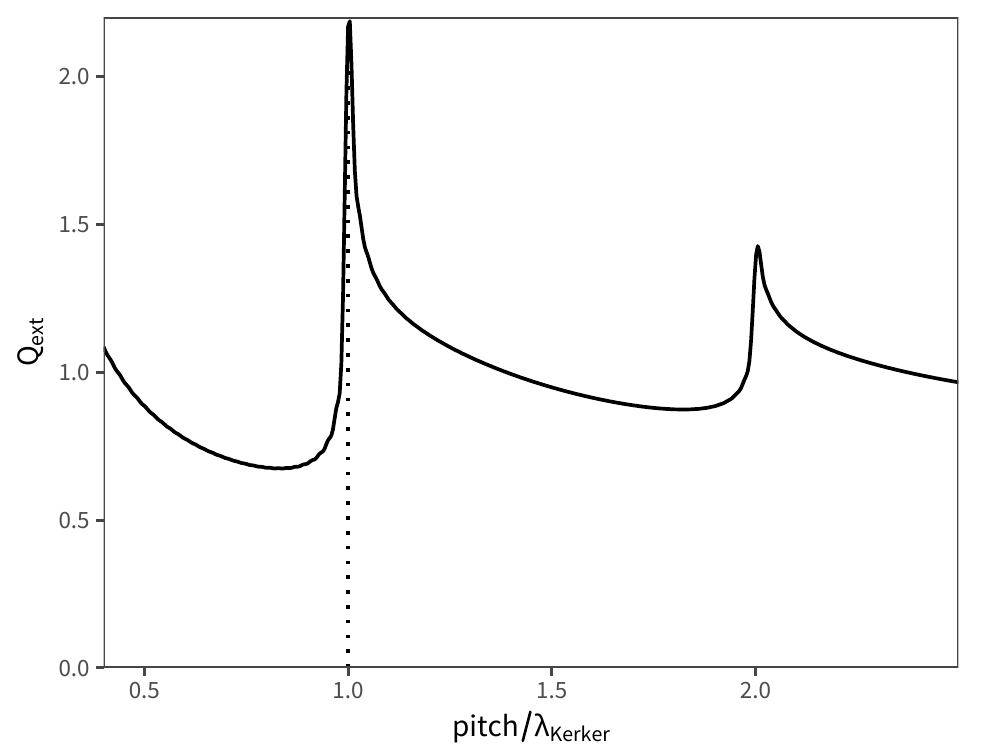}
    \caption{Extinction efficiency at $\lambda = \lambda_\text{Kerker} = 1208$\,nm for a chain of $N = 50$ silicon nanospheres as a function of $\text{pitch}/\lambda_\text{Kerker}$. The chain is illuminated with circular polarisation at normal incidence.}
	\label{figS6}
\end{suppfigure}  

\clearpage
\section{GoldHelix cluster description}

The following listing contains the cartesian positions of the 320 Au nanospheres used in the GoldHelix model of Fig.~4 of the main text. 

\begin{Verbatim}[fontsize=\tiny]
material x y z radius
Au  1.732050808e+01  1.000000000e+01 -2.139583333e+02 5
Au  1.000000000e+01  1.732050808e+01 -2.085416667e+02 5
Au  1.224646799e-15  2.000000000e+01 -2.031250000e+02 5
Au -1.000000000e+01  1.732050808e+01 -1.977083333e+02 5
Au -1.732050808e+01  1.000000000e+01 -1.922916667e+02 5
Au -2.000000000e+01  2.449293598e-15 -1.868750000e+02 5
Au -1.732050808e+01 -1.000000000e+01 -1.814583333e+02 5
Au -1.000000000e+01 -1.732050808e+01 -1.760416667e+02 5
Au -3.673940397e-15 -2.000000000e+01 -1.706250000e+02 5
Au  1.000000000e+01 -1.732050808e+01 -1.652083333e+02 5
Au  1.732050808e+01 -1.000000000e+01 -1.597916667e+02 5
Au  2.000000000e+01 -4.898587197e-15 -1.543750000e+02 5
Au  1.732050808e+01  1.000000000e+01 -1.489583333e+02 5
Au  1.000000000e+01  1.732050808e+01 -1.435416667e+02 5
Au  2.388680239e-14  2.000000000e+01 -1.381250000e+02 5
Au -1.000000000e+01  1.732050808e+01 -1.327083333e+02 5
Au -1.732050808e+01  1.000000000e+01 -1.272916667e+02 5
Au -2.000000000e+01  7.347880795e-15 -1.218750000e+02 5
Au -1.732050808e+01 -1.000000000e+01 -1.164583333e+02 5
Au -1.000000000e+01 -1.732050808e+01 -1.110416667e+02 5
Au -8.572527594e-15 -2.000000000e+01 -1.056250000e+02 5
Au  1.000000000e+01 -1.732050808e+01 -1.002083333e+02 5
Au  1.732050808e+01 -1.000000000e+01 -9.479166667e+01 5
Au  2.000000000e+01 -9.797174393e-15 -8.937500000e+01 5
Au  1.732050808e+01  1.000000000e+01 -8.395833333e+01 5
Au  1.000000000e+01  1.732050808e+01 -7.854166667e+01 5
Au  4.654895798e-14  2.000000000e+01 -7.312500000e+01 5
Au -1.000000000e+01  1.732050808e+01 -6.770833333e+01 5
Au -1.732050808e+01  1.000000000e+01 -6.229166667e+01 5
Au -2.000000000e+01  4.777360478e-14 -5.687500000e+01 5
Au -1.732050808e+01 -1.000000000e+01 -5.145833333e+01 5
Au -1.000000000e+01 -1.732050808e+01 -4.604166667e+01 5
Au -4.899825158e-14 -2.000000000e+01 -4.062500000e+01 5
Au  1.000000000e+01 -1.732050808e+01 -3.520833333e+01 5
Au  1.732050808e+01 -1.000000000e+01 -2.979166667e+01 5
Au  2.000000000e+01 -1.469576159e-14 -2.437500000e+01 5
Au  1.732050808e+01  1.000000000e+01 -1.895833333e+01 5
Au  1.000000000e+01  1.732050808e+01 -1.354166667e+01 5
Au  5.144754518e-14  2.000000000e+01 -8.125000000e+00 5
Au -1.000000000e+01  1.732050808e+01 -2.708333333e+00 5
Au -1.732050808e+01  1.000000000e+01  2.708333333e+00 5
Au -2.000000000e+01  1.714505519e-14  8.125000000e+00 5
Au -1.732050808e+01 -1.000000000e+01  1.354166667e+01 5
Au -1.000000000e+01 -1.732050808e+01  1.895833333e+01 5
Au -5.389683878e-14 -2.000000000e+01  2.437500000e+01 5
Au  1.000000000e+01 -1.732050808e+01  2.979166667e+01 5
Au  1.732050808e+01 -1.000000000e+01  3.520833333e+01 5
Au  2.000000000e+01 -1.959434879e-14  4.062500000e+01 5
Au 1.732050808e+01 1.000000000e+01 4.604166667e+01 5
Au 1.000000000e+01 1.732050808e+01 5.145833333e+01 5
Au 5.634613237e-14 2.000000000e+01 5.687500000e+01 5
Au -1.000000000e+01  1.732050808e+01  6.229166667e+01 5
Au -1.732050808e+01  1.000000000e+01  6.770833333e+01 5
Au -2.000000000e+01  9.309791596e-14  7.312500000e+01 5
Au -1.732050808e+01 -1.000000000e+01  7.854166667e+01 5
Au -1.000000000e+01 -1.732050808e+01  8.395833333e+01 5
Au -5.879542597e-14 -2.000000000e+01  8.937500000e+01 5
Au  1.000000000e+01 -1.732050808e+01  9.479166667e+01 5
Au  1.732050808e+01 -1.000000000e+01  1.002083333e+02 5
Au  2.000000000e+01 -9.554720956e-14  1.056250000e+02 5
Au 1.732050808e+01 1.000000000e+01 1.110416667e+02 5
Au 1.000000000e+01 1.732050808e+01 1.164583333e+02 5
Au 1.322989931e-13 2.000000000e+01 1.218750000e+02 5
Au -1.000000000e+01  1.732050808e+01  1.272916667e+02 5
Au -1.732050808e+01  1.000000000e+01  1.327083333e+02 5
Au -2.000000000e+01  9.799650316e-14  1.381250000e+02 5
Au -1.732050808e+01 -1.000000000e+01  1.435416667e+02 5
Au -1.000000000e+01 -1.732050808e+01  1.489583333e+02 5
Au -6.369401317e-14 -2.000000000e+01  1.543750000e+02 5
Au  1.000000000e+01 -1.732050808e+01  1.597916667e+02 5
Au  1.732050808e+01 -1.000000000e+01  1.652083333e+02 5
Au  2.000000000e+01 -2.939152318e-14  1.706250000e+02 5
Au 1.732050808e+01 1.000000000e+01 1.760416667e+02 5
Au 1.000000000e+01 1.732050808e+01 1.814583333e+02 5
Au 1.371975803e-13 2.000000000e+01 1.868750000e+02 5
Au -1.000000000e+01  1.732050808e+01  1.922916667e+02 5
Au -1.732050808e+01  1.000000000e+01  1.977083333e+02 5
Au -2.000000000e+01  1.028950904e-13  2.031250000e+02 5
Au -1.732050808e+01 -1.000000000e+01  2.085416667e+02 5
Au -1.000000000e+01 -1.732050808e+01  2.139583333e+02 5
Au  1.914252577e+01  5.793419313e+00 -2.026997755e+02 5
Au  1.368120395e+01  1.458851118e+01 -1.972831088e+02 5
Au  4.554014583e+00  1.947462326e+01 -1.918664421e+02 5
Au -5.793419313e+00  1.914252577e+01 -1.864497755e+02 5
Au -1.458851118e+01  1.368120395e+01 -1.810331088e+02 5
Au -1.947462326e+01  4.554014583e+00 -1.756164421e+02 5
Au -1.914252577e+01 -5.793419313e+00 -1.701997755e+02 5
Au -1.368120395e+01 -1.458851118e+01 -1.647831088e+02 5
Au -4.554014583e+00 -1.947462326e+01 -1.593664421e+02 5
Au  5.793419313e+00 -1.914252577e+01 -1.539497755e+02 5
Au  1.458851118e+01 -1.368120395e+01 -1.485331088e+02 5
Au  1.947462326e+01 -4.554014583e+00 -1.431164421e+02 5
Au  1.914252577e+01  5.793419313e+00 -1.376997755e+02 5
Au  1.368120395e+01  1.458851118e+01 -1.322831088e+02 5
Au  4.554014583e+00  1.947462326e+01 -1.268664421e+02 5
Au -5.793419313e+00  1.914252577e+01 -1.214497755e+02 5
Au -1.458851118e+01  1.368120395e+01 -1.160331088e+02 5
Au -1.947462326e+01  4.554014583e+00 -1.106164421e+02 5
Au -1.914252577e+01 -5.793419313e+00 -1.051997755e+02 5
Au -1.368120395e+01 -1.458851118e+01 -9.978310881e+01 5
Au -4.554014583e+00 -1.947462326e+01 -9.436644214e+01 5
Au  5.793419313e+00 -1.914252577e+01 -8.894977547e+01 5
Au  1.458851118e+01 -1.368120395e+01 -8.353310881e+01 5
Au  1.947462326e+01 -4.554014583e+00 -7.811644214e+01 5
Au  1.914252577e+01  5.793419313e+00 -7.269977547e+01 5
Au  1.368120395e+01  1.458851118e+01 -6.728310881e+01 5
Au  4.554014583e+00  1.947462326e+01 -6.186644214e+01 5
Au -5.793419313e+00  1.914252577e+01 -5.644977547e+01 5
Au -1.458851118e+01  1.368120395e+01 -5.103310881e+01 5
Au -1.947462326e+01  4.554014583e+00 -4.561644214e+01 5
Au -1.914252577e+01 -5.793419313e+00 -4.019977547e+01 5
Au -1.368120395e+01 -1.458851118e+01 -3.478310881e+01 5
Au -4.554014583e+00 -1.947462326e+01 -2.936644214e+01 5
Au  5.793419313e+00 -1.914252577e+01 -2.394977547e+01 5
Au  1.458851118e+01 -1.368120395e+01 -1.853310881e+01 5
Au  1.947462326e+01 -4.554014583e+00 -1.311644214e+01 5
Au  1.914252577e+01  5.793419313e+00 -7.699775475e+00 5
Au  1.368120395e+01  1.458851118e+01 -2.283108808e+00 5
Au 4.554014583e+00 1.947462326e+01 3.133557859e+00 5
Au -5.793419313e+00  1.914252577e+01  8.550224525e+00 5
Au -1.458851118e+01  1.368120395e+01  1.396689119e+01 5
Au -1.947462326e+01  4.554014583e+00  1.938355786e+01 5
Au -1.914252577e+01 -5.793419313e+00  2.480022453e+01 5
Au -1.368120395e+01 -1.458851118e+01  3.021689119e+01 5
Au -4.554014583e+00 -1.947462326e+01  3.563355786e+01 5
Au  5.793419313e+00 -1.914252577e+01  4.105022453e+01 5
Au  1.458851118e+01 -1.368120395e+01  4.646689119e+01 5
Au  1.947462326e+01 -4.554014583e+00  5.188355786e+01 5
Au 1.914252577e+01 5.793419313e+00 5.730022453e+01 5
Au 1.368120395e+01 1.458851118e+01 6.271689119e+01 5
Au 4.554014583e+00 1.947462326e+01 6.813355786e+01 5
Au -5.793419313e+00  1.914252577e+01  7.355022453e+01 5
Au -1.458851118e+01  1.368120395e+01  7.896689119e+01 5
Au -1.947462326e+01  4.554014583e+00  8.438355786e+01 5
Au -1.914252577e+01 -5.793419313e+00  8.980022453e+01 5
Au -1.368120395e+01 -1.458851118e+01  9.521689119e+01 5
Au -4.554014583e+00 -1.947462326e+01  1.006335579e+02 5
Au  5.793419313e+00 -1.914252577e+01  1.060502245e+02 5
Au  1.458851118e+01 -1.368120395e+01  1.114668912e+02 5
Au  1.947462326e+01 -4.554014583e+00  1.168835579e+02 5
Au 1.914252577e+01 5.793419313e+00 1.223002245e+02 5
Au 1.368120395e+01 1.458851118e+01 1.277168912e+02 5
Au 4.554014583e+00 1.947462326e+01 1.331335579e+02 5
Au -5.793419313e+00  1.914252577e+01  1.385502245e+02 5
Au -1.458851118e+01  1.368120395e+01  1.439668912e+02 5
Au -1.947462326e+01  4.554014583e+00  1.493835579e+02 5
Au -1.914252577e+01 -5.793419313e+00  1.548002245e+02 5
Au -1.368120395e+01 -1.458851118e+01  1.602168912e+02 5
Au -4.554014583e+00 -1.947462326e+01  1.656335579e+02 5
Au  5.793419313e+00 -1.914252577e+01  1.710502245e+02 5
Au  1.458851118e+01 -1.368120395e+01  1.764668912e+02 5
Au  1.947462326e+01 -4.554014583e+00  1.818835579e+02 5
Au 1.914252577e+01 5.793419313e+00 1.873002245e+02 5
Au 1.368120395e+01 1.458851118e+01 1.927168912e+02 5
Au 4.554014583e+00 1.947462326e+01 1.981335579e+02 5
Au -5.793419313e+00  1.914252577e+01  2.035502245e+02 5
Au -1.458851118e+01  1.368120395e+01  2.089668912e+02 5
Au -1.947462326e+01  4.554014583e+00  2.143835579e+02 5
Au -1.914252577e+01 -5.793419313e+00  2.198002245e+02 5
Au -1.368120395e+01 -1.458851118e+01  2.252168912e+02 5
Au  1.995883968e+01  1.282465851e+00 -1.914412176e+02 5
Au  1.664362927e+01  1.109006785e+01 -1.860245509e+02 5
Au  8.868771833e+00  1.792609512e+01 -1.806078843e+02 5
Au -1.282465851e+00  1.995883968e+01 -1.751912176e+02 5
Au -1.109006785e+01  1.664362927e+01 -1.697745509e+02 5
Au -1.792609512e+01  8.868771833e+00 -1.643578843e+02 5
Au -1.995883968e+01 -1.282465851e+00 -1.589412176e+02 5
Au -1.664362927e+01 -1.109006785e+01 -1.535245509e+02 5
Au -8.868771833e+00 -1.792609512e+01 -1.481078843e+02 5
Au  1.282465851e+00 -1.995883968e+01 -1.426912176e+02 5
Au  1.109006785e+01 -1.664362927e+01 -1.372745509e+02 5
Au  1.792609512e+01 -8.868771833e+00 -1.318578843e+02 5
Au  1.995883968e+01  1.282465851e+00 -1.264412176e+02 5
Au  1.664362927e+01  1.109006785e+01 -1.210245509e+02 5
Au  8.868771833e+00  1.792609512e+01 -1.156078843e+02 5
Au -1.282465851e+00  1.995883968e+01 -1.101912176e+02 5
Au -1.109006785e+01  1.664362927e+01 -1.047745509e+02 5
Au -1.792609512e+01  8.868771833e+00 -9.935788428e+01 5
Au -1.995883968e+01 -1.282465851e+00 -9.394121762e+01 5
Au -1.664362927e+01 -1.109006785e+01 -8.852455095e+01 5
Au -8.868771833e+00 -1.792609512e+01 -8.310788428e+01 5
Au  1.282465851e+00 -1.995883968e+01 -7.769121762e+01 5
Au  1.109006785e+01 -1.664362927e+01 -7.227455095e+01 5
Au  1.792609512e+01 -8.868771833e+00 -6.685788428e+01 5
Au  1.995883968e+01  1.282465851e+00 -6.144121762e+01 5
Au  1.664362927e+01  1.109006785e+01 -5.602455095e+01 5
Au  8.868771833e+00  1.792609512e+01 -5.060788428e+01 5
Au -1.282465851e+00  1.995883968e+01 -4.519121762e+01 5
Au -1.109006785e+01  1.664362927e+01 -3.977455095e+01 5
Au -1.792609512e+01  8.868771833e+00 -3.435788428e+01 5
Au -1.995883968e+01 -1.282465851e+00 -2.894121762e+01 5
Au -1.664362927e+01 -1.109006785e+01 -2.352455095e+01 5
Au -8.868771833e+00 -1.792609512e+01 -1.810788428e+01 5
Au  1.282465851e+00 -1.995883968e+01 -1.269121762e+01 5
Au  1.109006785e+01 -1.664362927e+01 -7.274550949e+00 5
Au  1.792609512e+01 -8.868771833e+00 -1.857884283e+00 5
Au 1.995883968e+01 1.282465851e+00 3.558782384e+00 5
Au 1.664362927e+01 1.109006785e+01 8.975449051e+00 5
Au 8.868771833e+00 1.792609512e+01 1.439211572e+01 5
Au -1.282465851e+00  1.995883968e+01  1.980878238e+01 5
Au -1.109006785e+01  1.664362927e+01  2.522544905e+01 5
Au -1.792609512e+01  8.868771833e+00  3.064211572e+01 5
Au -1.995883968e+01 -1.282465851e+00  3.605878238e+01 5
Au -1.664362927e+01 -1.109006785e+01  4.147544905e+01 5
Au -8.868771833e+00 -1.792609512e+01  4.689211572e+01 5
Au  1.282465851e+00 -1.995883968e+01  5.230878238e+01 5
Au  1.109006785e+01 -1.664362927e+01  5.772544905e+01 5
Au  1.792609512e+01 -8.868771833e+00  6.314211572e+01 5
Au 1.995883968e+01 1.282465851e+00 6.855878238e+01 5
Au 1.664362927e+01 1.109006785e+01 7.397544905e+01 5
Au 8.868771833e+00 1.792609512e+01 7.939211572e+01 5
Au -1.282465851e+00  1.995883968e+01  8.480878238e+01 5
Au -1.109006785e+01  1.664362927e+01  9.022544905e+01 5
Au -1.792609512e+01  8.868771833e+00  9.564211572e+01 5
Au -1.995883968e+01 -1.282465851e+00  1.010587824e+02 5
Au -1.664362927e+01 -1.109006785e+01  1.064754491e+02 5
Au -8.868771833e+00 -1.792609512e+01  1.118921157e+02 5
Au  1.282465851e+00 -1.995883968e+01  1.173087824e+02 5
Au  1.109006785e+01 -1.664362927e+01  1.227254491e+02 5
Au  1.792609512e+01 -8.868771833e+00  1.281421157e+02 5
Au 1.995883968e+01 1.282465851e+00 1.335587824e+02 5
Au 1.664362927e+01 1.109006785e+01 1.389754491e+02 5
Au 8.868771833e+00 1.792609512e+01 1.443921157e+02 5
Au -1.282465851e+00  1.995883968e+01  1.498087824e+02 5
Au -1.109006785e+01  1.664362927e+01  1.552254491e+02 5
Au -1.792609512e+01  8.868771833e+00  1.606421157e+02 5
Au -1.995883968e+01 -1.282465851e+00  1.660587824e+02 5
Au -1.664362927e+01 -1.109006785e+01  1.714754491e+02 5
Au -8.868771833e+00 -1.792609512e+01  1.768921157e+02 5
Au  1.282465851e+00 -1.995883968e+01  1.823087824e+02 5
Au  1.109006785e+01 -1.664362927e+01  1.877254491e+02 5
Au  1.792609512e+01 -8.868771833e+00  1.931421157e+02 5
Au 1.995883968e+01 1.282465851e+00 1.985587824e+02 5
Au 1.664362927e+01 1.109006785e+01 2.039754491e+02 5
Au 8.868771833e+00 1.792609512e+01 2.093921157e+02 5
Au -1.282465851e+00  1.995883968e+01  2.148087824e+02 5
Au -1.109006785e+01  1.664362927e+01  2.202254491e+02 5
Au -1.792609512e+01  8.868771833e+00  2.256421157e+02 5
Au -1.995883968e+01 -1.282465851e+00  2.310587824e+02 5
Au -1.664362927e+01 -1.109006785e+01  2.364754491e+02 5
Au  1.972656258e+01 -3.295865383e+00 -1.801826598e+02 5
Au  1.873163702e+01  7.008978143e+00 -1.747659931e+02 5
Au  1.271758444e+01  1.543577164e+01 -1.693493264e+02 5
Au  3.295865383e+00  1.972656258e+01 -1.639326598e+02 5
Au -7.008978143e+00  1.873163702e+01 -1.585159931e+02 5
Au -1.543577164e+01  1.271758444e+01 -1.530993264e+02 5
Au -1.972656258e+01  3.295865383e+00 -1.476826598e+02 5
Au -1.873163702e+01 -7.008978143e+00 -1.422659931e+02 5
Au -1.271758444e+01 -1.543577164e+01 -1.368493264e+02 5
Au -3.295865383e+00 -1.972656258e+01 -1.314326598e+02 5
Au  7.008978143e+00 -1.873163702e+01 -1.260159931e+02 5
Au  1.543577164e+01 -1.271758444e+01 -1.205993264e+02 5
Au  1.972656258e+01 -3.295865383e+00 -1.151826598e+02 5
Au  1.873163702e+01  7.008978143e+00 -1.097659931e+02 5
Au  1.271758444e+01  1.543577164e+01 -1.043493264e+02 5
Au  3.295865383e+00  1.972656258e+01 -9.893265976e+01 5
Au -7.008978143e+00  1.873163702e+01 -9.351599309e+01 5
Au -1.543577164e+01  1.271758444e+01 -8.809932642e+01 5
Au -1.972656258e+01  3.295865383e+00 -8.268265976e+01 5
Au -1.873163702e+01 -7.008978143e+00 -7.726599309e+01 5
Au -1.271758444e+01 -1.543577164e+01 -7.184932642e+01 5
Au -3.295865383e+00 -1.972656258e+01 -6.643265976e+01 5
Au  7.008978143e+00 -1.873163702e+01 -6.101599309e+01 5
Au  1.543577164e+01 -1.271758444e+01 -5.559932642e+01 5
Au  1.972656258e+01 -3.295865383e+00 -5.018265976e+01 5
Au  1.873163702e+01  7.008978143e+00 -4.476599309e+01 5
Au  1.271758444e+01  1.543577164e+01 -3.934932642e+01 5
Au  3.295865383e+00  1.972656258e+01 -3.393265976e+01 5
Au -7.008978143e+00  1.873163702e+01 -2.851599309e+01 5
Au -1.543577164e+01  1.271758444e+01 -2.309932642e+01 5
Au -1.972656258e+01  3.295865383e+00 -1.768265976e+01 5
Au -1.873163702e+01 -7.008978143e+00 -1.226599309e+01 5
Au -1.271758444e+01 -1.543577164e+01 -6.849326424e+00 5
Au -3.295865383e+00 -1.972656258e+01 -1.432659758e+00 5
Au  7.008978143e+00 -1.873163702e+01  3.984006909e+00 5
Au  1.543577164e+01 -1.271758444e+01  9.400673576e+00 5
Au  1.972656258e+01 -3.295865383e+00  1.481734024e+01 5
Au 1.873163702e+01 7.008978143e+00 2.023400691e+01 5
Au 1.271758444e+01 1.543577164e+01 2.565067358e+01 5
Au 3.295865383e+00 1.972656258e+01 3.106734024e+01 5
Au -7.008978143e+00  1.873163702e+01  3.648400691e+01 5
Au -1.543577164e+01  1.271758444e+01  4.190067358e+01 5
Au -1.972656258e+01  3.295865383e+00  4.731734024e+01 5
Au -1.873163702e+01 -7.008978143e+00  5.273400691e+01 5
Au -1.271758444e+01 -1.543577164e+01  5.815067358e+01 5
Au -3.295865383e+00 -1.972656258e+01  6.356734024e+01 5
Au  7.008978143e+00 -1.873163702e+01  6.898400691e+01 5
Au  1.543577164e+01 -1.271758444e+01  7.440067358e+01 5
Au  1.972656258e+01 -3.295865383e+00  7.981734024e+01 5
Au 1.873163702e+01 7.008978143e+00 8.523400691e+01 5
Au 1.271758444e+01 1.543577164e+01 9.065067358e+01 5
Au 3.295865383e+00 1.972656258e+01 9.606734024e+01 5
Au -7.008978143e+00  1.873163702e+01  1.014840069e+02 5
Au -1.543577164e+01  1.271758444e+01  1.069006736e+02 5
Au -1.972656258e+01  3.295865383e+00  1.123173402e+02 5
Au -1.873163702e+01 -7.008978143e+00  1.177340069e+02 5
Au -1.271758444e+01 -1.543577164e+01  1.231506736e+02 5
Au -3.295865383e+00 -1.972656258e+01  1.285673402e+02 5
Au  7.008978143e+00 -1.873163702e+01  1.339840069e+02 5
Au  1.543577164e+01 -1.271758444e+01  1.394006736e+02 5
Au  1.972656258e+01 -3.295865383e+00  1.448173402e+02 5
Au 1.873163702e+01 7.008978143e+00 1.502340069e+02 5
Au 1.271758444e+01 1.543577164e+01 1.556506736e+02 5
Au 3.295865383e+00 1.972656258e+01 1.610673402e+02 5
Au -7.008978143e+00  1.873163702e+01  1.664840069e+02 5
Au -1.543577164e+01  1.271758444e+01  1.719006736e+02 5
Au -1.972656258e+01  3.295865383e+00  1.773173402e+02 5
Au -1.873163702e+01 -7.008978143e+00  1.827340069e+02 5
Au -1.271758444e+01 -1.543577164e+01  1.881506736e+02 5
Au -3.295865383e+00 -1.972656258e+01  1.935673402e+02 5
Au  7.008978143e+00 -1.873163702e+01  1.989840069e+02 5
Au  1.543577164e+01 -1.271758444e+01  2.044006736e+02 5
Au  1.972656258e+01 -3.295865383e+00  2.098173402e+02 5
Au 1.873163702e+01 7.008978143e+00 2.152340069e+02 5
Au 1.271758444e+01 1.543577164e+01 2.206506736e+02 5
Au 3.295865383e+00 1.972656258e+01 2.260673402e+02 5
Au -7.008978143e+00  1.873163702e+01  2.314840069e+02 5
Au -1.543577164e+01  1.271758444e+01  2.369006736e+02 5
Au -1.972656258e+01  3.295865383e+00  2.423173402e+02 5
Au -1.873163702e+01 -7.008978143e+00  2.477340069e+02 5 
\end{Verbatim}

\bibliography{references} 
\bibliographystyle{rsc} 